\documentclass[twocolumn]{jetc20132e}
\usepackage{amsmath,graphicx}
\usepackage{amssymb}
\usepackage[normalem]{ulem}
\usepackage{color}

\def\re#1{(\ref{#1})}   
\def\Fo{\textit{Fo}}

\title{ 
Experimental aspects of heat conduction beyond Fourier
}

\author{P. V\'an$^{123}$, B. Cz\'el$^2$, T. F\"ul\"op$^{123}$, Gy. Gr\'of$^2$, \'A. Gyenis$^2$  and J. Verh\'as$^3$ 
   \affiliation{
    $^1$Dept. of Theoretical Physics, Institute for Particle and Nuclear Physics, Wigner Research Centre for Physics, HAS,  \\
    H-1525 Budapest, P.O.Box 49, Hungary; \\
    $^2$Dept. of Energy Engineering, Budapest Univ. of Technology and Economics,\\
  H-1111, Budapest, M\H uegyetem rkp. 3-9,  Hungary;\\
$^3$Montavid Thermodynamic Research Group \\
Email: van.peter@wigner.mta.hu
}}

\date{
{\small 
Heat conduction experiments are performed in order to identify effects beyond Fourier. Two experimental setups are discussed. First, a simple experiment by a heterogeneous material is investigated from the point of view of generalized heat conduction, then the classical laser flash method is analysed. 
}}

\begin{document}

\maketitle

\section{Introduction}

The theory of heat conduction is the parade ground of testing and developing generalized thermodynamic theories \cite{JosPre89a,JosPre90a,Cim09a}. Recently, a linear irreversible thermodynamic framework of heat conduction  was introduced, where the deviation from local equilibrium is characterized by a single internal variable and by the generalization of the entropy current density via a current intensity factor \cite{Nyi91a1,Van01a2,CimVan05a}. A general constitutive evolution equation of the current density of internal energy was derived via introducing linear relationship between the thermodynamic forces and fluxes. The Fourier, Maxwell-Cattaneo-Vernotte, Guyer-Krumhansl, Jeffreys type and Green-Naghdi type equations of heat conduction were obtained as special cases \cite{VanFul12a}. This constitutive equation incorporates memory effects and weak nonlocality at the same time, however, only a local entropy function is assumed, that does not depend on the space derivatives of the internal energy, the basic 
state variable.

The balance of internal energy is written as
\begin{equation}
 \rho \frac{\partial e}{\partial t}+ \partial^i q^i = 0, 
 \label{ebal} \end{equation}
where $\rho$ is the density, $e$ is the specific internal energy and $q^i$ is the conductive current density of the internal energy, the heat flux. For the internal energy we assume a constant specific heat $c$ in the equation of state $e=c T$, where $T$ is the temperature. The above evolution equation is written in a substantial form, assuming negligible production of internal energy. $\partial_t$ denotes the substantial time derivative, the partial time derivative of the corresponding scalar quantity on the material manifold \cite{FulVan12a}. The heat flux is interpreted accordingly. The space derivative $\partial^i$ is used for the gradient in the material framework.

Then one may introduce two kind of irreversibilities. A vectorial internal variable together with the assumption that the heat is not parallel to the entropy current density, $j^0_s= B^{ij}q_j$, leads to the following nonlocal, relaxation type constitutive evolution equation of heat flux $q^i$ is obtained in the following form \cite{VanFul12a}:
\begin{eqnarray}
\tau\frac{\partial}{\partial t} q^i + q^i &=& \lambda_1 \partial^i \frac{1}{T} +
     \lambda_2 \frac{\partial}{\partial t}\left(\partial^i \frac{1}{T}\right) + 
     a_1 \partial^{ij} q^j + a_2 \partial^{jj} q^i + \nonumber\\ 
     & & b_1  \frac{\partial}{\partial t}(\partial^{ij}q^j)  + 
     b_2  \frac{\partial}{\partial t}(\partial^{jj}q^i). 
\label{geneq}\end{eqnarray}

The material parameters $\tau$, $\lambda_1$, $\lambda_2$, $a_1$, $a_2$, $b_1$, $b_2$ are nonnegative and not independent, because
\begin{equation}
a_1 \lambda_2 = b_1 \lambda_1, \qquad a_2 \lambda_2 = b_2 \lambda_1. 
\label{relcoeff}\end{equation}

Based on these theoretical considerations, the universality of the theory was demonstrated by showing that various heat conduction mechanisms and material structures lead to the above form of the constitutive relation. For example, material heterogeneity, with the possibility of two temperatures, is one such possible substructure. Therefore, \re{geneq} is interpreted as a universal, effective approach to heat conduction beyond Fourier. 

This rendering leads to a new point of view for the experimental investigations. Phonon propagation is not the only possibility of non-Fourier heat conduction, but new, mesoscopic structural effects can also play a role. The question is whether and under what conditions we can observe these deviations. One may think, and can also partially show, that in \re{geneq} the additional terms have the effect of driving the solution toward the solution of the Fourier equation. Therefore, suppressing the dissipation -- as it was performed in the classical experiments of phonon based low temperature heat conduction (see e.g. \cite{JacWal71a}) -- is not an option.

We show some results of the analysis of two experimental setups in order to identify possible deviations from the Fourier heat conduction: 

\begin{enumerate}
 \item \textit{Simple heterogeneous materials}. In these experiments, a layered periodical heterogeneous structure is the subject of abrupt temperature jump at one of the boundaries. The heat transfer properties of the layers (paper and air) are different. The additional material parameters of the equation are determined by the experiments, fitting the solutions of different models of heat conduction. 
 
 \item \textit{Heat conduction measurements with flash method.} Here, we analyse models of ordinary flash experiments from the point of view of beyond-Fourier heat conduction. Some benchmarks are established for the parameters of the material, the device and the operation for the identification of non-Fourier effects. 
\end{enumerate}

\section{Book experiment}

\begin{figure}
\begin{center}
       \includegraphics[width=0.45\textwidth]{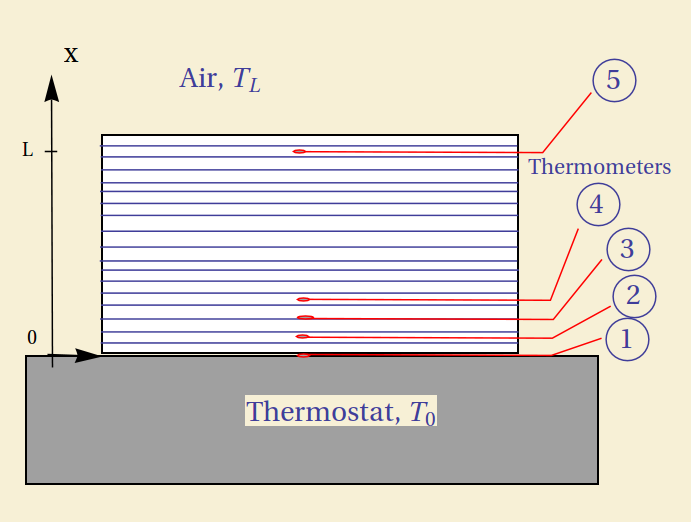}
\end{center}
\caption{ \label{expfig}
Experimental setup. The position of the thermometers is indicated by the numbers.}
\end{figure}

In this simple measurement we constructed a layered structure of 200 paper sheets, initially at ambient temperature $T_L= 29.9 ^\circ C$. At the beginning of the measurement, the structure has been contacted to a thermostat, while measuring the temperature at 5 different points of the structure. The experimental setup is sketched in Fig \ref{expfig}. The sample contained 200 
layers of $19 cm \times 15 cm$ sheets that were fastened at one of the shorter sides. The total thickness of the sample was $L=20mm$. The pages were $0.09mm$ thick and the air between them initially $0.01mm$, an estimation based on the difference of the total and compressed thickness. The density of the paper is $800 kg/m^3$ and the isobaric specific heat $1340 J/kgK$. The temperature was measured by a K-type thermometer at the thermostat and by copper-constantan thermocouple wires with diameter $d=0.1mm$ between the papers. The first thermometer was built in at the copper plate surface of the thermostat, and the second, third and fourth thermometers were $1 mm$, $2 mm$ and $3 mm$ distant from the surface of the sample at the side of the thermostat, respectively. The position of the fifth thermometer was $19mm$ from the thermostat surface. The thermometers were positioned $7.5cm$ from the free, not fastened sides.

\begin{figure}
\begin{center}
       \includegraphics[width=0.45\textwidth]{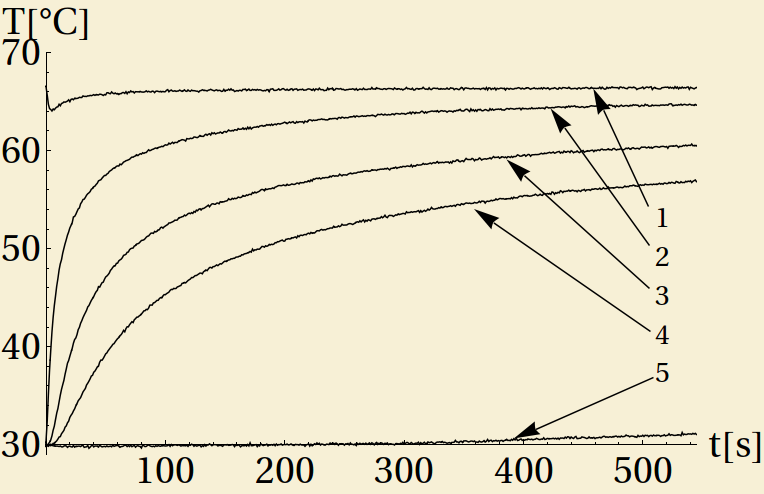}
\end{center}
\caption{ \label{datfig}
Temperatures of the thermometers 1-5, the serial numbers are increasing downward.}
\end{figure}

The measured temperatures are shown in Fig \ref{datfig} as a function of time. One can see that the thermostat cannot be considered homogeneous at the beginning, the temperature of the first thermometer drops by some centigrades in the first seconds of the measurement. The temperature of the farthest thermometer, number 5, increases only a few centigrades during the measurement. 
We have introduced two effective models for the evaluation of the data. In both cases, the thickness of the thermocouple wires was neglected and the problem was considered as one dimensional, the $x$ axis being perpendicular to the surface of the layers. The thermometers were comoving with the material, therefore the heat conduction models were interpreted in a material, i.e. first Piola-Kirchhoff framework.

\subsection{Effective nonlinear Fourier equation}

Here, we introduced the  Fourier equation with linearly temperature dependent heat conductivity, 
\begin{equation}
q =  -\lambda_F [1+ b (T-T_{ref})] \frac{\partial T}{\partial x}. 
\label{nlFou}\end{equation}
where $q$ is the $x$ component of the heat flux and the space derivative is a material one at the $x$ direction. $\lambda_F$ is the Fourier heat conduction coefficient at $T_{ref}=0 ^\circ C$, and $b$ characterizes the linear temperature dependence of the thermal conductivity. 

The initial temperature of the sample was uniform $T(t=0,x)= T_L= 29.9 ^\circ C$. The boundary condition in the free surface was constant, at the ambient temperature $T(t,x=L) = T_L$. At the side of the thermostat we have assumed a constant thermal conductance $\alpha$, therefore, the corresponding boundary condition is 
$$
\lambda_F[1 + b (T(t, x=0)-T_{ref})] \frac{\partial T}{\partial x}(t,x=0) = \alpha [T(t,x=0)-T_0].
$$ 
The free parameters of the model were the heat conduction coefficient $\lambda_F$, the thermal conductance $\alpha$, the parameter $b$, and the temperature of the thermostat $T_0$. In order to determine the best effective model parameters we considered the measured temperature data of thermometers $2,3,4$ in every $5s$ to be fitted by the following partial differential equation, subject to the above boundary and initial conditions:
\begin{equation}
 \rho c \frac{\partial T}{\partial t} -
 \lambda_F\frac{\partial}{\partial x} \left([1 + b (T-T_{ref})] \frac{\partial T}{\partial x}\right) =0.
\end{equation}
The best fit parameters are the following:
\renewcommand{\arraystretch}{1.5}
\begin{center}
\begin{tabular}{c|c|c|c}
		& $\lambda_F [W/m\,K]$ & $b [1/K]$& $T_2 [^\circ C]$  \\ \hline
Values 		& $0.140$          & $-0.008$ & $69.65$ \\ \hline
Stand. err.    & $0.006$ 	    &  $0.0003$ & $0.15$  \\
    \end{tabular}\\
\vskip .21cm
{Table 1. Fitted parameters of the nonlinear Fourier model}\end{center}

\begin{figure}
\begin{center}
       \includegraphics[width=0.45\textwidth]{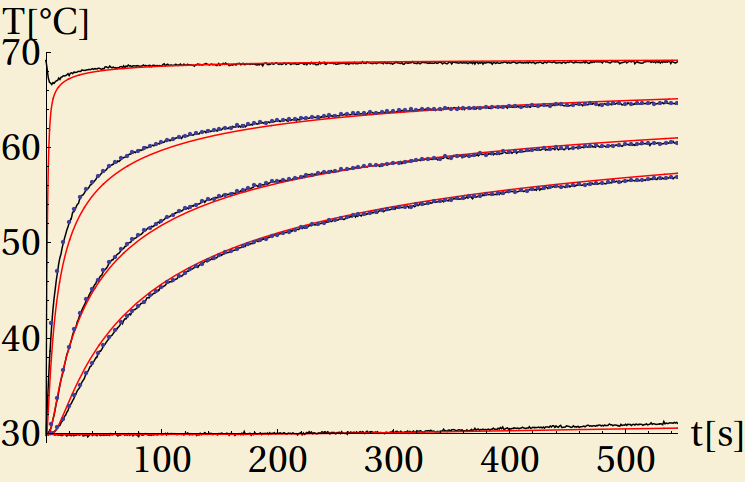}
\end{center}
\caption{ \label{nlFoufig}
Fitted nonlinear Fourier equation. Black, uncertain line: thermometer data, blue dots: data points of the fit, red smooth line: fitted model.}
\end{figure}

The calculations resulted in a high and uncertain value of the thermal conductance $\alpha$, indicating that the boundary can be considered at a constant temperature, and that the fit is not sensitive to this parameter. The negative $b$ may indicate the role of the weight on the top of the sheets. The goodness of  the fit can be characterised by $R^2 = 0.9992$. 
We have plotted the data and the fit together in Fig \ref{nlFoufig}. The red lines denote the fitted function, the blue dots indicate the data points used for the fit from thermometers 2, 3 and 4, and the black lines show the complete measurement data according to Fig \ref{datfig}.

\subsection{Effective Guyer-Krumhansl equation}

Our second model introduces the same number of fit parameters with a reduced version of the generalized constitutive heat conduction, where $\lambda_2=0$, therefore, $b_1=b_2=0$. The one dimensional form of \re{geneq} can be written as
\begin{equation}
\tau\frac{\partial}{\partial t} q +  q =
   -\lambda_F \frac{\partial T}{\partial x} + a \frac{\partial^2 q}{\partial x^2}.
\label{GK}\end{equation}

The Jeffreys type heat conduction could be a similar simplification, with a similar number of parameters. The role of the nonlocal terms with these boundary and initial conditions is probably similar \cite{VanFul12a}. 

In this case, the initial temperature distribution in the sample is the same as in the nonlinear Fourier case $T(t=0,x)= T_L=29.9 ^\circ C$ and the boundary conditions are similar: $T(t,x=L) = T_L$ and $q(t, x=0) = -\alpha_0 [T(t,x=0) - T_0]$. However, this model requires an additional initial and also a further boundary condition.  We assume that in case of uniform temperature distribution the initial heat flux was zero  $q(t=0,x)= 0$ and at the free side of the sample we assume constant thermal conductance, with a large constant coefficient $q(t, x=L) = -\alpha_L [T(t, L) - T_L]$ in order to ensure approximately constant boundary temperature. 
The thermal conductances were chosen $\alpha_0=\alpha_L= 50000 \pm 40000 W/m^2K$. The error estimate is based on sensitivity calculations, the heat conduction model is not sensitive in these parameters, similarly to the previous nonlinear Fourier one. The free parameters of the model are the heat conduction coefficient $\lambda_F$, the relaxation time $\tau$, the Guyer-Krumhansl parameter $a$ and the temperature of the thermostat $T_0$. 

In order to determine the best effective model parameters, we considered the same temperature data as in the case of the nonlinear Fourier model. The model introduces the following system of partial differential equations:
\begin{eqnarray}
  \rho c \frac{\partial T}{\partial t} + \frac{\partial q}{\partial x} &=& 0, \label{GKeqn}\\
  \tau\frac{\partial}{\partial t} q +  q &=&
   -\lambda_F \frac{\partial T}{\partial x} + a \frac{\partial^2 q}{\partial x^2}.   
\end{eqnarray}

The best fit parameters are the following:
\renewcommand{\arraystretch}{1.5}
\begin{center}
\begin{tabular}{c|c|c|c|c}
		& $\lambda\, [W/m\,K]$ & $\tau\, [s]$ & $a\, [m^2]$             & $T_0\, [^\circ C]$  \\ \hline
Values 		& $0.05243$        & $194.9$    & $1.415\times 10^{-5}$ & $69.52$ \\ \hline
Stand. err.     & $0.00003$ 	   & $0.1$      & $9\times 10^{-8}$     & $0.05$  \\
    \end{tabular}\\
\vskip .21cm
{Table 2. Fitted parameters of the Guyer-Krumhansl model.}
\end{center}

\begin{figure}
\begin{center}
       \includegraphics[width=0.45\textwidth]{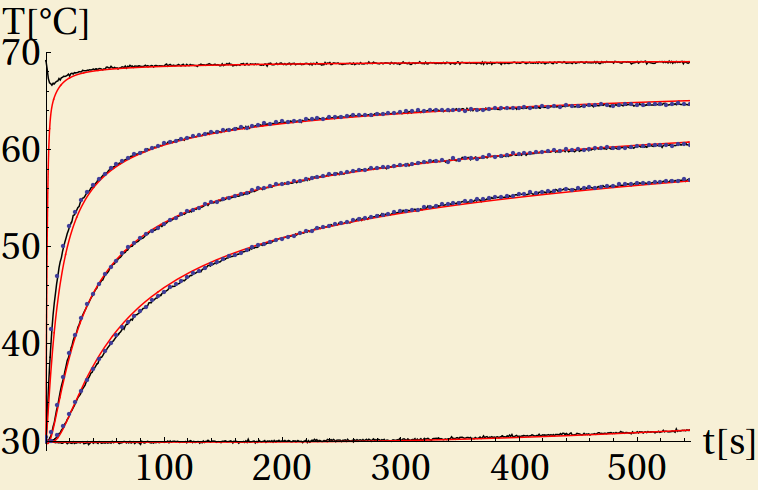}
\end{center}
\caption{ \label{GKfig}
Fitted Guyer-Krumhansl equation. Black, uncertain line: thermometer data, blue dots: data points of the fit, red smooth line: fitted model.}
\end{figure}

The goodness of  the fit can be characterised by $R^2 = 0.99997$. We have plotted the data and the fit together in Fig \ref{GKfig}. The red lines denote the fitted function, the blue dots indicate the data points used for the fit from thermometers 2, 3 and 4, and the black lines show the complete measurement data according to Fig \ref{datfig}.

The Guyer-Krumhansl equation seems to fit the data slightly better than the nonlinear Fourier model. The thermostat temperatures are approximately the same. However, neither fits are perfect, there are visible deviations. Remarkable is the large difference of the Fourier heat conduction coefficients in case of the different models. The heat conduction of paper is $\lambda_{\rm paper} =0.11-0.13\,W/m\,K$, and that of air is $\lambda_{\rm air} =0.024\, W/m\,K$. In our evaluation, the temperature dependent heat conduction coefficient is $\lambda_F(T=T_L) = 0.11\,W/m\,K$ at air temperature and $\lambda_F(T=T_0) = 0.063\,W/m\,K$ at the termostat.

\section{Flash simulations}

The laser flash method is a common method to measure the thermal diffusivity ($D$, see its definition below) of solid materials having medium and high thermal conductivity ($\lambda_F$). During the measurement, a small disk-shaped specimen (with $1$-$3 mm$ thickness and $10$-$30 mm$ diameter) is subjected to a short and high intensity laser pulse on the front face, and the temperature response is recorded on the rear face. $D$ is then calculated based on the time to reach half of the maximum temperature rise of the rear face. The method was originally proposed by Parker et al. \cite{ParEta61a} in 1961, then further improved by applying corrections for finite pulse time \cite{CapLeh63a} and heat loss effects \cite{Cow63a,ClaTay75a}. Since the 1960s, numerous extensions of the original method have been introduced including e.g., the application for liquids, heterogeneous materials, and two- or three-layered specimens. Recently, the ultrafast laser flash method has been proposed by Baba et al. \cite{BabEta11a} 
for 
the measurement of thin films attached 
to a substrate. The thickness of the film can be less than $100nm$, for which the laser pulse duration should be in the order of magnitude of picoseconds. In this parameter domain, the application of hyperbolic heat conduction models might be necessary. (It should be noted that the ultrafast laser flash method has significant differences in the way of temperature measurement compared to the classical laser flash method.) In the present study the Maxwell-Cattaneo-Vernotte (MCV) type heat conduction model was applied to simulate laser flash experiments. Our aim was to find the parameters for which the relaxation effect of the MCV equation can be observed via the measurement.

\begin{figure}
\begin{center}
       \includegraphics[width=0.45\textwidth]{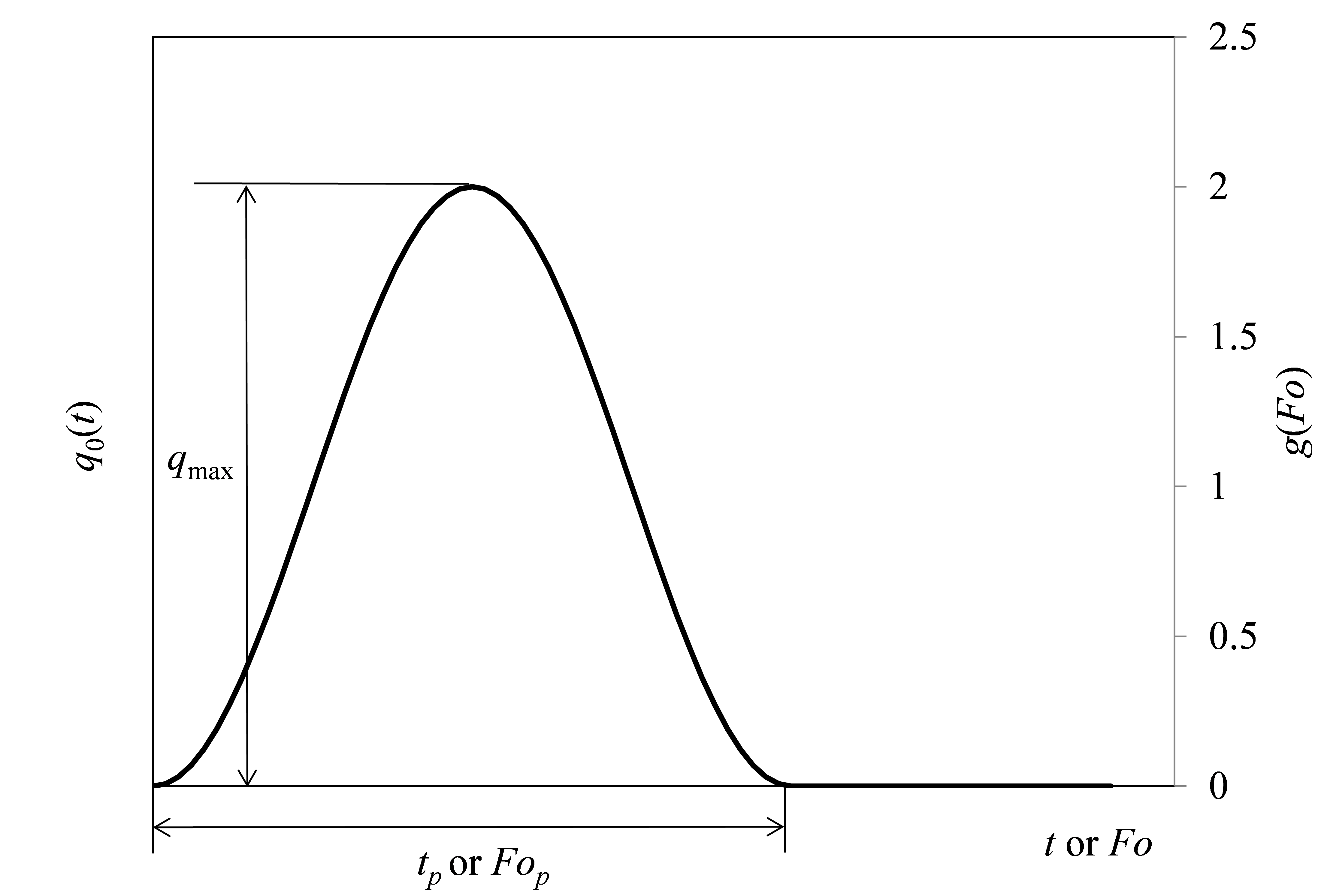}
\end{center}
\caption{The shape of the applied pulse. \label{pulsefig}
}
\end{figure}

The laser flash experiment was simulated for a single layer specimen that is solid, homogeneous and material properties are constant. One-dimensional heat conduction through the thickness of the specimen was assumed; heat losses were neglected. Heat conduction was modeled according to the MCV equation:
\begin{equation}
\tau\frac{\partial}{\partial t} q +  q =
   \lambda_F \frac{\partial T}{\partial x},
\label{MCV}\end{equation}
which is identical to \re{GK} when $a=0$. With the help of \re{ebal}, one can obtain the following partial differential equation for the temperature: 
\begin{equation}
 \tau \frac{\partial^2 T}{\partial t^2} + \frac{\partial T}{\partial t} = D  \frac{\partial^2 T}{\partial x^2},
\label{telegr}\end{equation}
where $D=\frac{\lambda_F}{\rho c}$ is the thermal diffusivity. We solve the differential equation with the following initial  conditions $T(t=0,x)=T_0$, $\frac{\partial T}{\partial t}(t=0,x) = 0$. At the front boundary, $x=0$, a heat pulse is introduced, with the following form (Fig \ref{pulsefig}):
\begin{equation}
q_0(t) = \left\{\begin{matrix}
                  \frac{1}{2} q_{max} \left[1-\cos(2\pi t/t_p)\right], & 0<t\leq t_p, \\
                  0, & t>t_p.
                 \end{matrix}
 \right.
\label{pulsefun}\end{equation}
The front boundary condition itself can be given with the help of the MCV equation \re{MCV}
\begin{equation}
\tau\frac{d}{d t} q_0 +  q_0 =
   -\lambda_F \frac{\partial T}{\partial x}(t,x=0).
\label{bMCV}\end{equation}
At the rear face, the boundary condition is required to be $\frac{\partial T}{\partial x}(t,x=L) = 0$.

In order to find the parameters, for which the relaxation effect of the MCV equation can be observed, we introduced the following dimensionless variables and parameters: 
\begin{eqnarray}
 \theta &=& \frac{T-T_0}{T_L-T_0}, \quad \text{where} \quad T_L= T_0+\frac{1}{\rho c L}\int_0^{t_p} q_0(t)dt,\\
 \Fo    &=& \frac{D t}{L^2},\\
 \xi   &=& \frac{x}{L}, \quad \text{therefore} \quad 0\leq \xi\leq 1,\\
 \Pi^2 &=& \frac{D \tau}{L^2}, \\
 g(\Fo) &=& \frac{q_0(Fo)}{\bar{q}}, \quad  \text{where} \quad \bar{q} = \frac{1}{\Fo_p} \int_0^{\Fo_p} q_0(\Fo)d\Fo.
\end{eqnarray}
Then we obtain the differential equation \re{telegr}  in a dimensionless form as follows:
\begin{equation}
 \Pi^2 \frac{\partial^2 \theta}{\partial \Fo^2} + \frac{\partial \theta}{\partial \Fo} =  \frac{\partial^2 \theta}{\partial \xi^2},
\label{telegrnd}\end{equation}
The initial conditions are $\theta(\Fo=0,\xi)=0$ and $\frac{\partial\theta}{\partial \Fo}(\Fo=0,\xi)=0$. The heat pulse is 
\begin{equation}
g(\Fo) = \left\{\begin{matrix}
                  1- \cos\left(2\pi \frac{\Fo}{\Fo_p}\right), & \ 0<\Fo \leq \Fo_p, \\
                  0, & \Fo>\Fo_p.
                 \end{matrix}
\right.
\label{pulsefunnd}\end{equation}
and the front boundary condition 
\begin{equation}
\Pi^2 \frac{d}{d t} g +  g =
   -\Fo_p \frac{\partial \theta}{\partial \xi}(\Fo,\xi=0).
\label{bMCVnd}\end{equation}
The rear face boundary condition is $\frac{\partial \theta}{\partial\xi}(\Fo,\xi=1)=0$.

The problem was solved numerically, using a simple explicit finite difference method. The key for obtaining a stable numerical solution was to keep the Courant number, $\textit{Cou} = \frac{\delta t}{\delta x}\sqrt{\frac{D}{\tau}}=\frac{\delta\Fo}{\delta\xi \Pi}$, close to but less than $1$. ($\textit{Cou}=0.99$ was applied.) 

The shape of the laser pulse $q(t)$ is shown in Fig \ref{pulsefig}. This function was chosen because it starts with $0$ derivative, its derivative is finite everywhere, and it has finite height and length. (Starting with non-zero derivative and/or a jump in the function can cause instability in the numerical solution.)
Looking at the dimensionless equations we can conclude that the problem can be fully characterized by two dimensionless parameters: $\Pi^2$ and $\Fo_p$. In a laser flash experiment, the rear face temperature history, $\theta(\Fo,\xi=1)$, is recorded. Hence, the effect of these two parameters on the rear face temperature history is to  be examined. 

\begin{figure}
\begin{center}
       \includegraphics[width=0.45\textwidth]{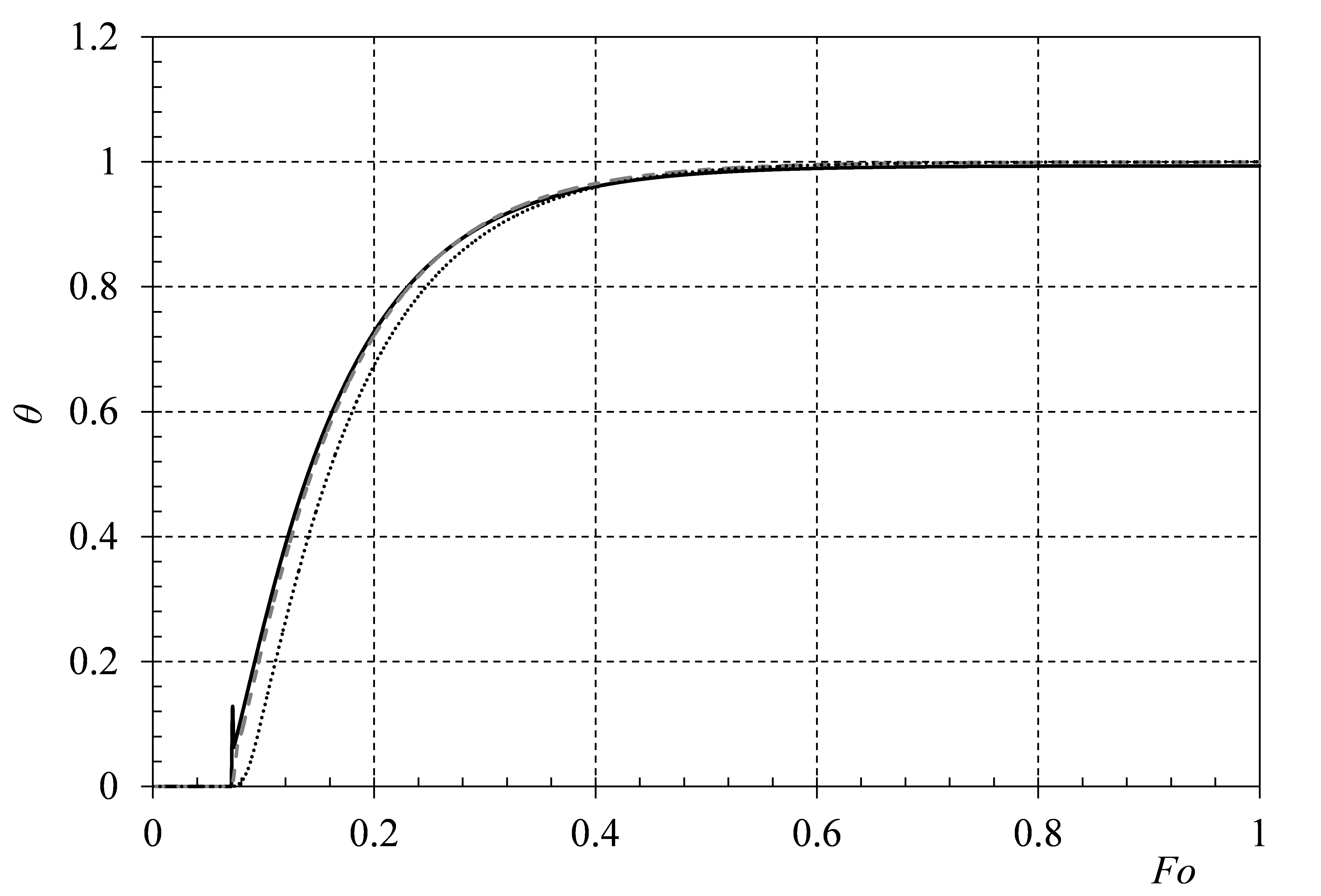}
\end{center}
\caption{Rear side temperature history, $\Pi^2=0.005$. Black solid lines – $\Fo_p =0.5\Pi^2$, grey dashed lines – $\Fo_p =2\Pi^2$, black dotted lines – $\Fo_p =8\Pi^2$ \label{ff1}
}
\end{figure}
\begin{figure}
\begin{center}
       \includegraphics[width=0.45\textwidth]{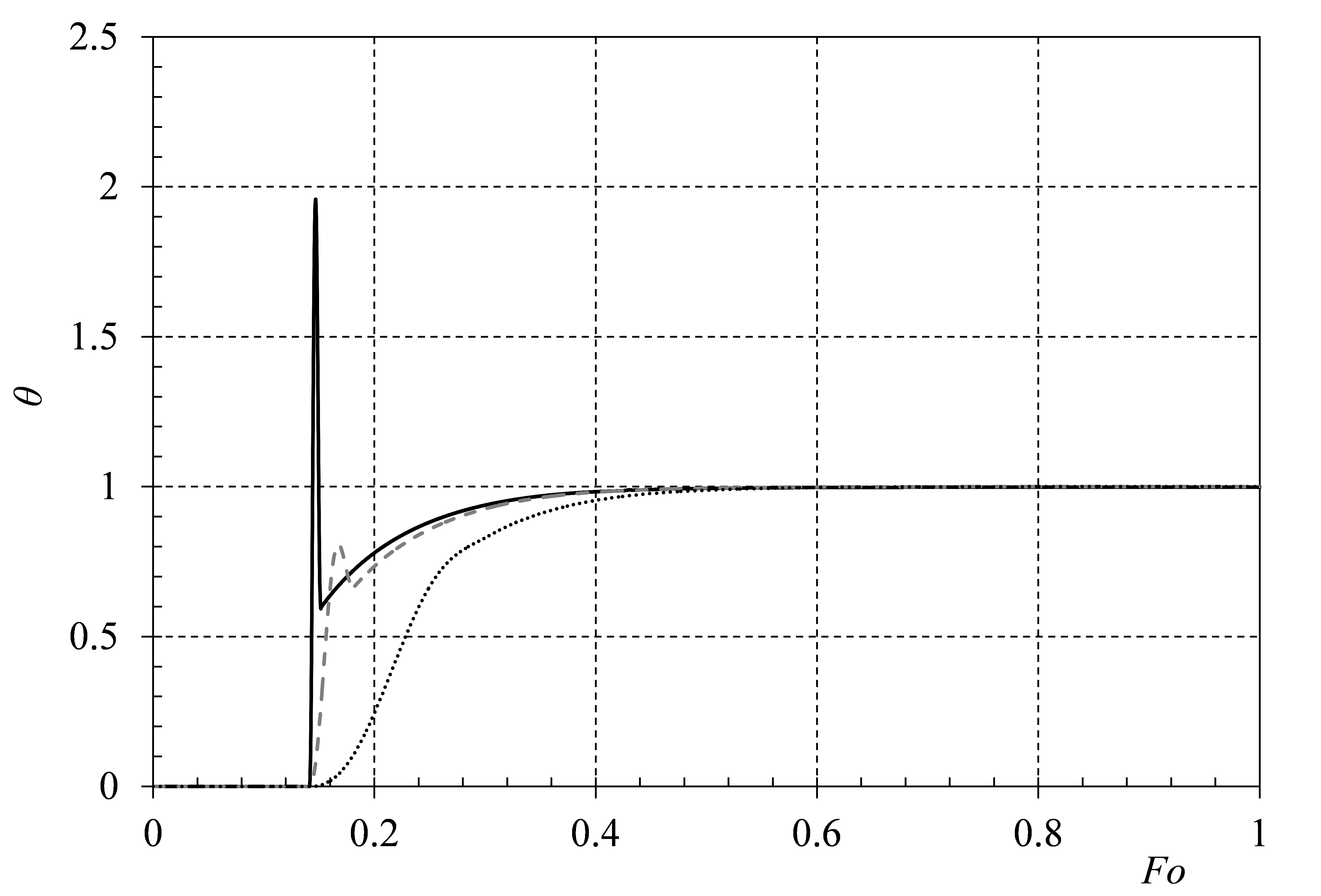}
\end{center}
\caption{Rear side temperature history, $\Pi^2=0.02$. Black solid lines – $\Fo_p =0.5\Pi^2$, grey dashed lines – $\Fo_p =2\Pi^2$, black dotted lines – $\Fo_p =8\Pi^2$ \label{ff2}
}
\end{figure}
\begin{figure}
\begin{center}
       \includegraphics[width=0.45\textwidth]{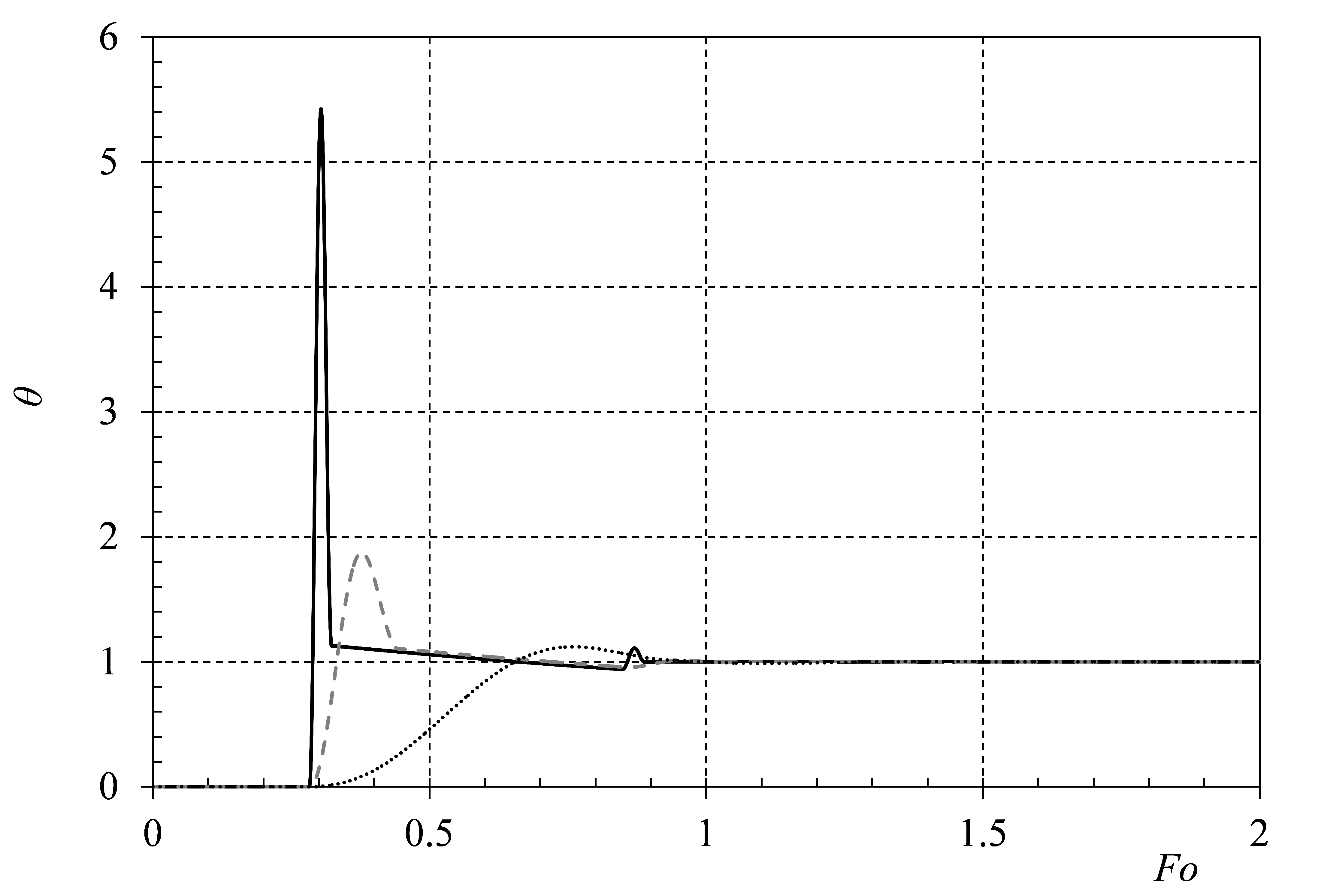}
\end{center}
\caption{Rear side temperature history, $\Pi^2=0.08$. Black solid lines – $\Fo_p =0.5\Pi^2$, grey dashed lines – $\Fo_p =2\Pi^2$, black dotted lines – $\Fo_p =8\Pi^2$ \label{ff3}
}
\end{figure}

Results of the numerical solution of the dimensionless problem are shown in Figs \ref{ff1}-\ref{ff3} for various values of $\Pi^2$ and $\Fo_p$. In Fig \ref{ff1}, where $\Pi^2=0.005$, only a slight relaxation effect can be seen for $\Fo_p=0.5\Pi^2$. For longer pulses, waves are not present in the temperature histories. The main difference in these temperature histories compared to the ones calculated using the classical Fourier model is that the temperature remains $0$ for a specified time before it starts increasing. In real measurements, it is difficult to detect this because of the noise in the temperature measurement, and applying a finite pulse time correction might appear to fix the deviance from the Fourier model. In Fig \ref{ff2}, where $\Pi^2=0.02$, a wave can be clearly seen in the temperature history for $\Fo_p=0.5\Pi^2$ and $\Fo_p=2\Pi^2$, which is a clear and easily detectable sign of the MCV equation. Increasing $\Pi^2$ further (see Fig \ref{ff3} where $\Pi^2=0.08$), a second wave can be seen 
in the temperature history for the shortest pulse. This means that the wave bounces back from the front side and reaches the rear side for the second time. The temperature histories for all three $\Fo_p$ values rise above the steady state value, which is again a detectable sign of the MCV equation.
The length of the pulse ($\Fo_p$) has a significant effect on the temperature history. When $\Fo_p$ is much less than $\Pi^2$ then the wave is very sharp, which causes high maximum temperature on the front side. When $\Fo_p$ is much more than $\Pi^2$ then the wave might disappear from the temperature history. This effect can be understood by comparing the two terms on the left hand side of equation  \re{bMCVnd}. We found that the optimal value of $\Fo_p$ is around $2 \Pi^2$, because in this case the two terms have the same order of magnitude.
\begin{figure}
\begin{center}
       \includegraphics[width=0.5\textwidth]{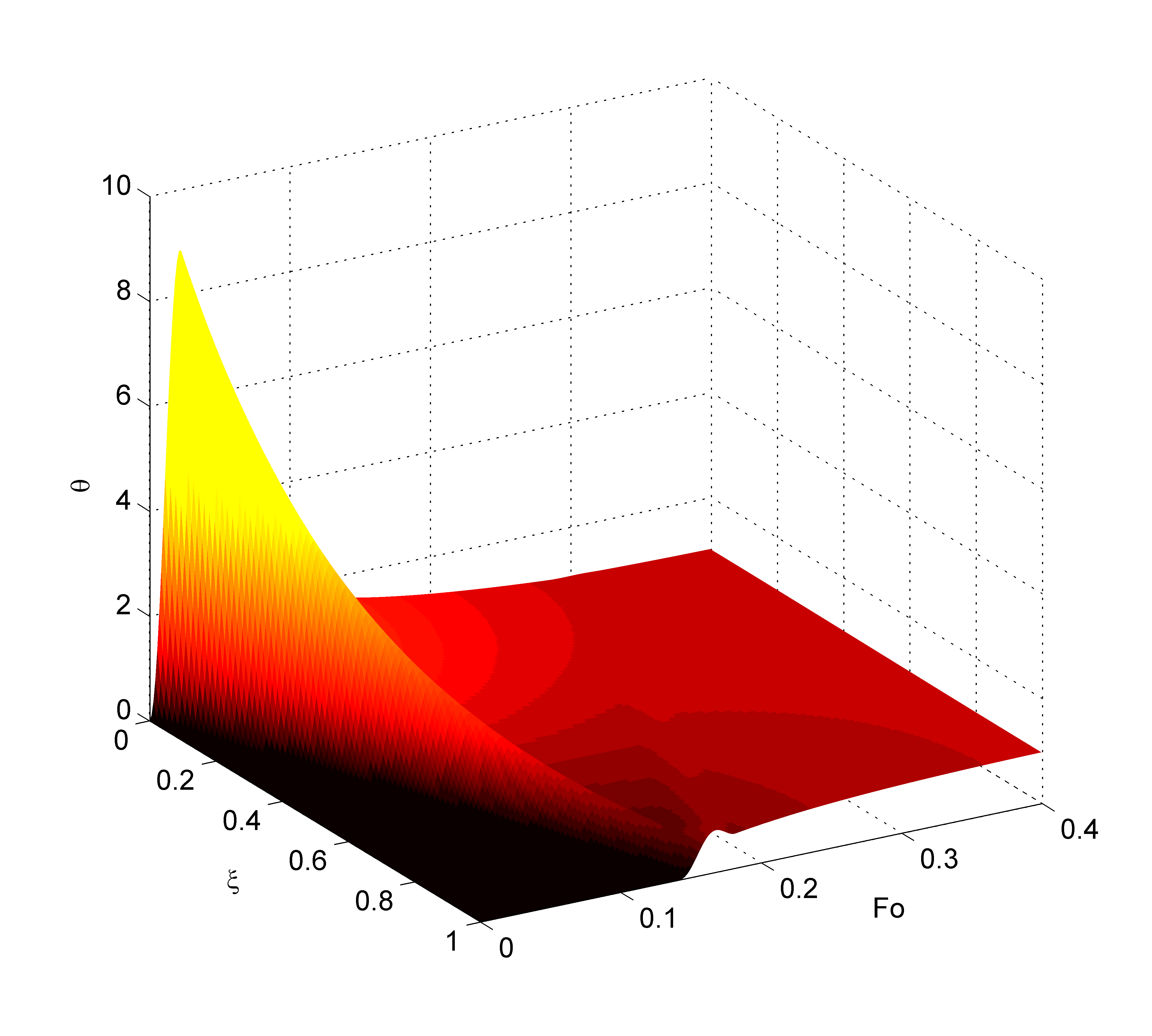}
\end{center}
\caption{Dimensionless temperature distribution for $\Pi^2=0.02$ and $\Fo_p=2\Pi^2$. \label{f3d}
}
\end{figure}

The value of $\Pi^2$ in case of the classical laser flash measurements is very low. In order to obtain higher values, special conditions are necessary. This is why it is desirable to find the smallest value of $\Pi^2$ for which the relaxation effect of the MCV equation can be reliably observed. Considering the results of the numerical calculations, we suggest $\Pi^2=0.02$ and $\Fo_p=2\Pi^2$ as target values for the detection of the relaxation effect with the laser flash method. The complete temperature distribution in time and space for this case is shown in Fig \ref{f3d}. The propagation of the wave caused by the laser pulse can be followed along the thickness. The wave bounces back from the rear side but does not reach the front side again. The maximum temperature on the front side is about 9 times higher than the steady state temperature.
\renewcommand{\arraystretch}{1.5}
\begin{center}
\begin{tabular}{c|c|c}
		& Al $@ 293K$ 		 & NaF  $@ 15K$        \\ \hline
$D\, [m^2/s]$	& $9.2\cdot 10^{-5}$     & $2.6$    		\\ \hline
$\tau\, [s]$	& $2.6\cdot 10^{-12}$ 	 & $6.8\cdot 10^{-7}$   \\ \hline
$L \, [m]$	& $1.1\cdot 10^{-7}$	 & $9.4\cdot 10^{-3}$    \\ \hline
$t_p\, [s]$	& $5.2\cdot 10^{-12}$    & $1.4\cdot 10^{-6}$     
    \end{tabular}\\
\vskip .21cm
{Table 3. Parameters according to  $\Pi^2=0.02$ and $\Fo_p=2\Pi^2$ for aluminium at $293K$ and for NaF at $15K$ (NaF properties are based on \cite{JacWal71a}.}
\end{center}

Finally, we calculated the recommended thickness ($L$) and pulse time ($t_p$) according to $\Pi^2=0.02$ and $\Fo_p=2\Pi^2$ for aluminum at $293K$ and NaF at $15K$. The results are summarized in Table 3. For aluminium at $293K$ the recommended thickness is about $110nm$, and the pulse time is about $5.2ps$. With this thickness, the measurement cannot be performed on a standalone sample, however, measurements on two-layer configurations might be possible \cite{BabEta11a}. This, naturally needs the modification of the mathematical model, and the target values of the dimensionless parameters must be revised. Regarding NaF at $15K$, the recommended thickness is $9.4mm$, and the pulse time is $1.4\mu s$. These  are close to the experimental values in \cite{JacWal71a} and not extreme, customised laser flash instruments are able to perform measurements with these parameters. The big challenge is to perform the measurement at $15K$. The second difficulty is that the thermal conductivity of NaF around $15K$ changes 
rapidly with temperature, which has to be considered in the mathematical model. Another option is to keep the temperature change caused by the pulse very small, which makes the temperature measurement very challenging. 

In conclusion, the detection of the relaxation effect of the MCV equation by the laser flash method in the classical configuration might be possible only for materials with extremely high conductivity and very sophisticated instrumentation. However, the possibility of detecting the relaxation effect seems to be open for a wider range of materials applying a modified multilayer configuration of the classical method like in \cite{BabEta11a}. For this the mathematical model must be modified to match the multilayer configuration, and the target values of the dimensionless parameters must be revised.

\section{Discussion}

The book experiment can be modeled reasonably well by both a nonlinear Fourier equation, with time dependent heat conductivity and by a linear Guyer-Krumhansl equation. \textit{These results are not conclusive, one cannot decide which model is better. 
} Especially, the high value of the relaxation time in the GK model should be handled by care. First, because there are uncertainties in the fitting, and in the modeling of the contact with the thermostat. Moreover, here the special qualitative phenomena of the Maxwell-Cattaneo-Vernotte equation may be suppressed by the higher order dissipation due to the last term of \re{GKeqn}. 

The analysis of the laser flash method highlighted the experimental parameters where qualitative effects due to the MCV equation can be identified. When dissipation beyond the Fourier equation cannot be suppressed, a similar analysis of the Guyer-Krumhansl and the Jeffreys type model is necessary.

\section{Acknowledgement}   
The work was supported by the grants OTKA K81161 and K104260.


\end{document}